\documentclass[twocolumn,aps,showpacs,nofootinbib,prd]{revtex4-1}
\usepackage{amssymb}
\usepackage{amsmath}
\usepackage{graphics}
\usepackage{epsfig}
\usepackage[dvips]{color}
\newcommand{\be}{\begin{equation}}
\newcommand{\ee}{\end{equation}}
\newcommand{\bea}{\begin{eqnarray}}
\newcommand{\eea}{\end{eqnarray}}
\newcommand{\beas}{\begin{eqnarray*}}
\newcommand{\eeas}{\end{eqnarray*}}

\newcommand{\nn}{\nonumber\\}

\newcommand{\bd}[1]{{\bf #1}}

\begin{document}
\title{Relating $\hat{q}$, $\eta/s$ and $\Delta E$ in an expanding Quark-Gluon Plasma}
\author{Alejandro
Ayala$^{1,5}$, Isabel Dominguez$^2$, Jamal Jalilian-Marian$^3$ and Maria Elena
  Tejeda-Yeomans$^{4}$}
  \address{
  $^1$Instituto de Ciencias
  Nucleares, Universidad Nacional Aut\'onoma de M\'exico, Apartado
  Postal 70-543, M\'exico Distrito Federal 04510,
  Mexico.\\
  $^2$Facultad de Ciencias F\'isico-Matem\'aticas, Universidad Aut\'onoma de Sinaloa,
  Avenida de las Am\'ericas y Boulevard Universitarios, Ciudad Universitaria,
  C.P. 80000, Culiac\'an, Sinaloa, M\'exico.\\
  $^3$Department of Natural Sciences, Baruch College, New York, NY, 10010, USA and\\
  CUNY Graduate Center, 365 Fifth Avenue, New York, NY 10016, USA.\\
  $^4$Departamento de F\'{\i}sica,
  Universidad de Sonora, Boulevard Luis Encinas J. y Rosales, Colonia
  Centro, Hermosillo, Sonora 83000, Mexico.\\
  $^5$Centre for Theoretical and Mathematical Physics, and Department of Physics,
  University of Cape Town, Rondebosch 7700, South Africa}

\begin{abstract}

We use linear viscous hydrodynamics to describe the energy and momentum deposited by a fast moving parton in a quark gluon plasma. This energy-momentum is in turn used to compute the probability density for the production of soft partons by means of the Cooper-Frye formula. We use this probability density to render manifest a relation between the average transverse momentum given to the fast moving parton from the medium $\hat{q}$, the entropy density to shear viscosity ratio $\eta/s$ and the energy lost by the fast moving parton $\Delta E$ in an expanding medium under similar conditions to those generated in nucleus-nucleus collisions at the LHC. We find that $\hat{q}$ increases linearly with $\Delta E$ for both trigger and away side partons that have been produced throughout the medium. On the other hand, $\eta/s$ is more stable with $\Delta E$. We also study how these transport coefficients vary with the geometrical location of the hard scattering that produces the fast moving partons. The behavior of $\hat{q}$, with $\Delta E$ is understood as arising from the length of medium the parton traverses from the point where it is produced. However, since  $\eta/s$ is proportional to the ratio of the length of medium traversed by the fast parton and the average number of scatterings it experiences, it has a milder dependence on the energy it loses. This study represents a tool to obtain a direct connection between transport coefficients and the description of in-medium energy loss within a linear viscous hydrodynamical evolution of the bulk.

\end{abstract}

\pacs{25.75.-q, 25.75.Gz, 12.38.Bx}
\maketitle

\section{Introduction}\label{I}

The results from experiments where heavy nuclei are collided at high energies, carried both at the BNL Relativistic Heavy-Ion Collider~\cite{RHIC} and the CERN Large Hadron Collider~\cite{LHC}, show that a state of matter, where quarks and gluons are not confined to individual nucleons is formed, the so called Quark-Gluon Plasma (QGP). The behavior of the QGP's soft bulk, that is particles with $p_T\lesssim 2$ GeV, can be accurately described using viscous hydrodynamics in the liquid regime~\cite{viscoshydro,Heinz}. One way to study the properties of the QGP is to consider the way that fast moving partons, either quarks or gluons, traveling through the QGP, transfer energy and momentum to the bulk. This energy-momentum is then converted into particles upon hadronization. The process is well described by hydrodynamics, where the source of energy and momentum is the fast moving parton~\cite{Casalderrey2, linhydro, Neufeld79}. 
In a gradient expansion to lowest non-trivial order, one can include the viscous effects to first order in the shear viscosity to entropy ratio $\eta/s$. This procedure is known as linear viscous hydrodynamics. Extracting the value of this transport coefficient for the QGP is at the core of current efforts both in the theoretical and experimental fronts~\cite{Jacki}.

An important ingredient for this description is the energy loss per unit length $dE/dx$ which enters as the coefficient describing the strength of the local hydrodynamic source term. Hadronization of the energy and momentum deposited into the medium can be carried out by means of the Cooper-Frye formula. In this way, linear viscous hydrodynamics provides a means to compute the probability density describing the production of soft-particles from the energy lost by a fast moving parton in the medium~\cite{previous}. 

Another important quantity that describes the interaction of fast moving partons and the QGP is the average transverse momentum squared per unit length transferred to the parton from the medium, the so called $\hat{q}$-parameter. This parameter represents the average broadening of the original fast parton's trajectory resulting from interactions induced by the medium. It has been suggested~\cite{previousqhat} that $\hat{q}$ for a thermal parton can be related to the shear viscosity to entropy ratio $\eta/s$ of the QGP plasma, since it is argued that $\hat{q}$ is a measure of the coupling strength of the medium. The relation suggested between these transport coefficients involves the temperature of the medium depending on the coupling strength regime, as
\begin{equation}
   \frac{T^3}{\hat{q}} \left\{
   \begin{array}{l} 
   \approx \frac{\eta }{s}~~, ~\mbox{weakly-coupled} \\
   \ll \frac{\eta }{s}~~, ~\mbox{strongly-coupled}.
   \end{array} \right.
   \label{qhat-old}
\end{equation}
Furthermore, there are collaborative efforts in the theoretical community to extract the jet transport parameter $\hat{q}$ using different approaches to energy-loss mechanisms for jet quenching at RHIC and LHC energies~\cite{qhatdata}. Further recent phenomenological studies in connection with underlying anomalies~\cite{qhat-a} and thermalization of minijets~\cite{qhat-deltaE} together with recent theoretical proposals using effective field theory to describe transport coefficients~\cite{qhat-quasi}, make evident the need for elucidating the interplay between these transport coefficients across the phenomenological landscape of the QGP.

Since both $\eta/s$ and $\hat{q}$ are transport coefficients describing the exchange of energy and momentum between fast partons and medium, a pertinent question is whether there is a quantitative relation between these parameters that can be extracted from the hydrodynamical picture. In this work we provide such relation using linear viscous hydrodynamics and the Cooper-Frye formula to describe the probability to produce soft partons from the energy-momentum deposited by the fast partons to the bulk. We show that a non-trivial relation exists for the case when the medium is considered as non-static. In doing so we also study the scaling of such transport coefficients with the parton's travelled length and their total energy lost while traversing the QGP. The paper is organized as follows: In Sec.~\ref{II} we derive a relation between the transport coefficients $\hat{q}$ and $\eta/s$ for a static medium. We show that in this case $\hat{q}$ is independent of $\eta/s$. In Sec.~\ref{III} we relax the condition of studying a static medium introducing a model for the energy loss that accounts for expansion. We show that under these conditions a non-trivial dependence between these transport coefficients emerges. We also study the effects that the position of the hard scattering and therefore of the path length travelled by the fast moving partons within the plasma, have on the extracted values for the transport coefficients. We show that these coefficients depend on the total energy lost and thus that triggering on events with a certain energy loss will yield different values of these parameters. We finally summarize and conclude in Sec.~\ref{IV}.

\section{$\hat{q}$ and $\eta/s$ in a static medium}\label{II}

The particle's multiplicity is given by the Cooper-Frye formula~\cite{Cooper-Frye}
\bea
   E\frac{dN}{d^3p}=\frac{1}{(2\pi )^3}\int d\Sigma_\mu p^\mu
   [f(p\cdot u) - f(p_0)],
   \label{Cooper-Frye}
\eea 
where $f(p\cdot u) - f(p_0)$ is the phase-space disturbance
produced by the fast moving parton on top of the equilibrium
distribution $f(p_0)$, with $\Sigma_\mu$ and $p_\mu$ representing the
freeze-out hypersurface and the particle's momentum,
respectively. The medium's total four-velocity $u^\mu \equiv u_0^\mu +
\delta u^\mu$ is made out of two parts: the background four-velocity
$u_0^\mu$ and the disturbance $\delta u^\mu$. This last contribution
is produced by the fast moving parton and can be computed using
linear viscous hydrodynamics once the source, representing the parton,
is specified. For a static background (which we assume for the moment) and in
the linear approximation, $u^\mu$ can be written as \bea u^\mu
&\equiv& u_0^\mu + \delta u^\mu
\nn\\ &=&\left(1,\frac{\mathbf{g}}{\epsilon_0(1+c_s^2)}\right),
   \label{fourvel}
\eea
where the spatial part of the medium's four-velocity,
${\mathbf{u}}={\mathbf{g}}/\epsilon_0(1+c_s^2)$, is written for
convenience in terms of the momentum density ${\mathbf{g}}$ associated
to the disturbance, with $\epsilon_0$ and $c_s$ the static
background's energy density and sound velocity, respectively. We focus
on events at central rapidity, $y\simeq 0$, and take the direction of
motion of the fast parton to be the $\hat{z}$ axis and the beam axis
to be the $\hat{x}$ axis. With this geometry, the transverse plane is
the $\hat{y}-\hat{z}$ plane and therefore, the momentum four-vector
for a (massless) particle is explicitly given by \bea p_\mu &=&
(E,p_x,p_y,p_z)\nn\\ &=& (p_T,0,p_T\sin\phi,p_T\cos\phi),
   \label{momcomponents}
\eea
where $\phi$ is the angle that the momentum vector ${\mathbf{p}}$
makes with the $\hat{z}$ axis. We use Bjorken's geometry thus,
\bea
   d^3p&=&p_Tdp_Td\phi dp_x\nn
   p_x&=&p_T\sinh y\nn
   dp_x&=&p_T\cosh y\  dy\nn
   E&=&p_T\cosh y
   \label{dist}
\eea
and therefore
\bea
    E\frac{dN}{d^3p}=\frac{dN}{p_Tdp_Td\phi dy}.
    \label{dist2}
\eea
For simplicity we consider a freeze-out hypersurface of constant time,
\bea
   d\Sigma_\mu=(d^3r,0,0,0).
   \label{hypersurface}
\eea
Therefore, using Eqs.~(\ref{Cooper-Frye}) and~(\ref{dist2}), the
particle momentum distribution around the direction of motion of a
fast moving parton is given by
\bea
   \frac{dN}{p_Tdp_Td\phi d^2r}=\frac{\Delta\tau(\Delta y)^2}{(2\pi )^3}
   p_T[f(p\cdot u) - f(p_0)],
   \label{distapprox}
\eea
with $\Delta\tau$ the freeze-out time interval, $d^2r$ the surface element in the transverse plane and where we have
assumed a perfect correlation between the space-time rapidity $\eta$ and
$y$ to substitute $\Delta\eta$ by $\Delta y$. We assume that the
equilibrium distribution is of the Boltzmann type. Assuming that the energy density and temperature are related
through Boltzmann's law
\bea
   \epsilon\propto T^4,
   \label{bolaw}
\eea
one gets
\bea
   \frac{\delta T}{T_0}=\frac{\delta\epsilon}{4\epsilon_0},
   \label{delbolaw}
\eea
where $T_0$ is the background medium's temperature and $\delta T$ is
the change in temperature caused by the passing of the fast
parton.

Since for the validity of linearized hydrodynamics, both
$\delta\epsilon$ and ${\mathbf{g}}$ need to be small quantities
compared to $\epsilon_0$, we can expand the difference $f(p\cdot u) -
f(p_0)$ to linear order
\bea 
f(p\cdot u) - f(p_0)&\simeq&
\left(\frac{p_T}{T_0}\right)\exp\left[ -p_T/T_0
  \right]\nn
  &\times&\left( \frac{\delta\epsilon}{4\epsilon_0} +
\frac{{\mathbf{g}}_y\sin\phi + {\mathbf{g}}_z\cos\phi}{\epsilon_0(1+c_s^2)} \right).
   \label{linearorder}
\eea

The energy and momentum densities, $\delta\epsilon$ and ${\mathbf g}_i$, can be found as the solution of the linearized viscous hydrodynamical equations once the source representing the fast moving parton is specified. The current associated to the source is given by 
\bea
   J^\nu(\mathbf{r},t)=\left(\frac{dE}{dx}\right)v^\nu\delta^3(\mathbf{x}-\mathbf{v}t),
\label{current}
\eea
where $(dE/dx)$ is the energy loss per unit length. $\delta\epsilon$ and ${\mathbf g}_i$ are given then by
\begin{eqnarray}
  \delta\epsilon &=&
   \left(\frac{1}{4\pi}\right)
   \left( \frac{d E}{d x} \right)\left(\frac{2v}{3\Gamma_s}\right)^2 
   \left(\frac{9}{8v}\right)I_{\delta\epsilon}(\alpha ,\beta )
\label{moddeltaeps}
\end{eqnarray}
and
\begin{eqnarray}
\bd{g}_i &=&
   \left(\frac{1}{4\pi}\right)
   \left( \frac{d E}{d x} \right)\left(\frac{2v}{3\Gamma_s}\right)^2 I_{\mathbf{g}_i}(\alpha ,\beta ),
\label{modgtlyz} 
\end{eqnarray}
where the integrals $I_{\delta\epsilon}$ and $I_{\mathbf{g}_i}$ are dimensionless functions representing the collected energy-momentum deposited by the source term when moving through the medium and are given in Refs.~\cite{ADY} in terms of dimensionless variables $\alpha $ and $\beta$ which encode the distance to the source, in units of the sound attenuation length
\begin{eqnarray}
\Gamma_s\equiv \frac{4 \eta }{3\epsilon_0(1+c_s^2)}.
\label{atlength}
\eea 
Therefore, the particle momentum distribution around the direction of
motion of a fast moving parton is given, in the linear approximation,
by
\bea
   \frac{dN}{p_Tdp_T d\phi d^2r}&=&\frac{\Delta\tau(\Delta y)^2}{(2\pi )^3}
   \frac{p_T^2}{T_0}\exp\left[ -p_T/T_0 \right]\nn
   &\times&
   \left( \frac{\delta\epsilon}{4\epsilon_0} + \frac{{\mathbf{g}}_y\sin\phi + {\mathbf{g}}_z\cos\phi}{\epsilon_0(1+c_s^2)} \right).
   \label{lineardist}
\eea

\begin{widetext}
\begin{figure*}[t]
\begin{center}
\includegraphics[scale=0.42]{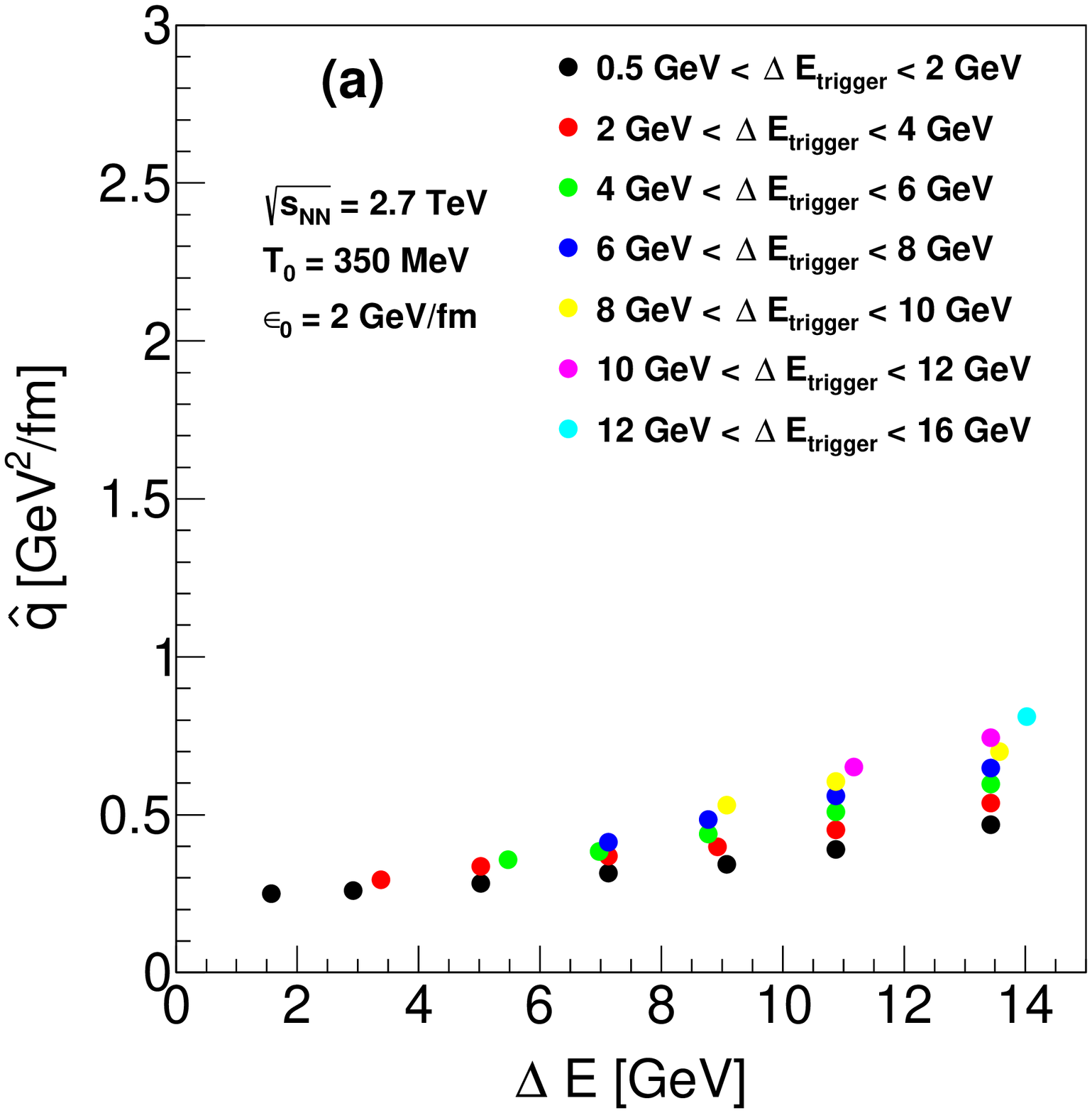}
\includegraphics[scale=0.42]{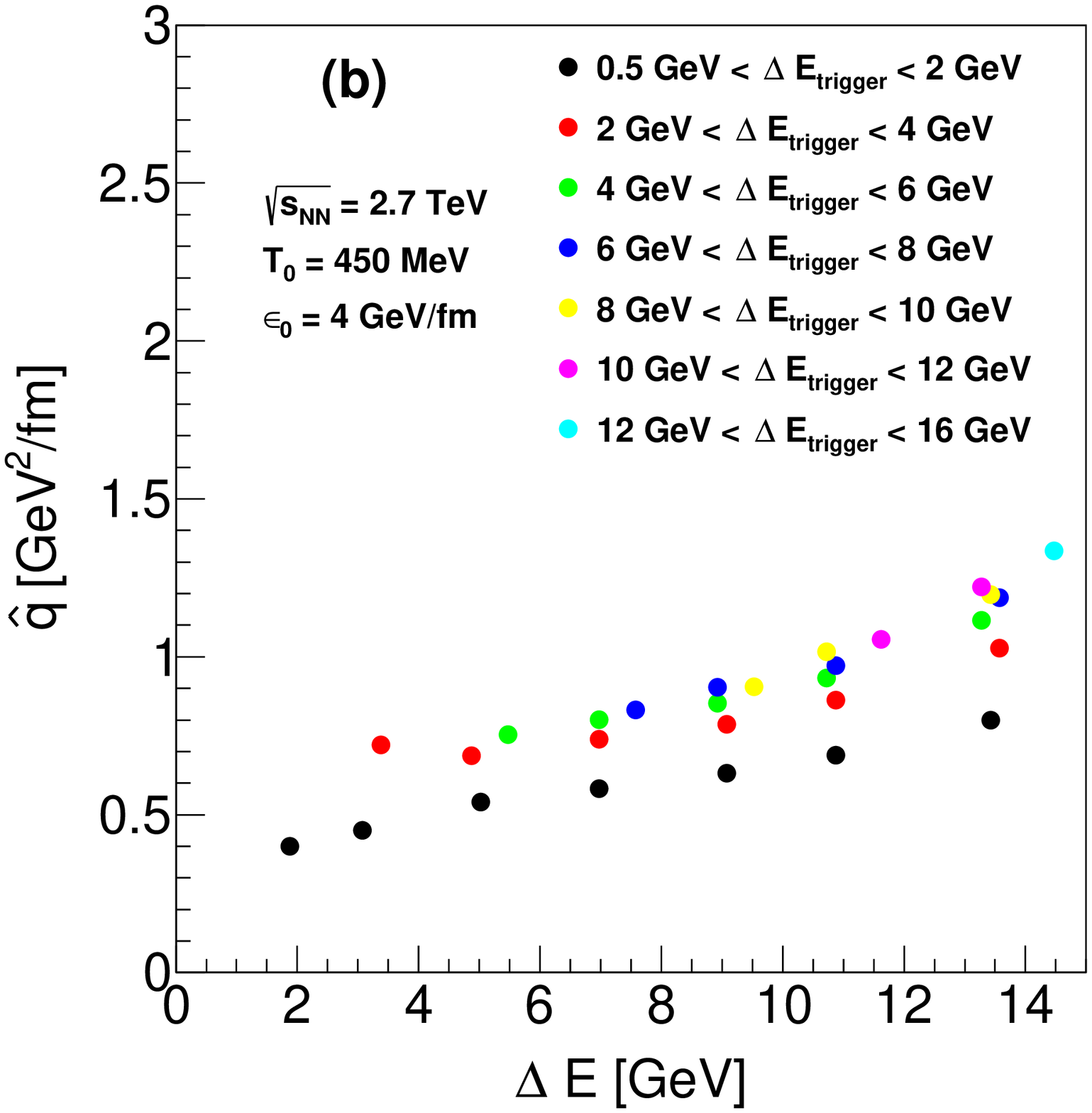}
\end{center}
\caption{$\hat{q}$ as a function of $\Delta E$ for the away-side particle for different trigger particle energy loses. The left (right) panel corresponds to a QGP temperature  $T_0=350$ ($T_0=450$) MeV and a corresponding $\epsilon_0=2$ ($\epsilon_0=4$) GeV/fm in the $1-d$ energy loss model.}
\label{fig1}
\end{figure*}
\end{widetext}

The probability density ${\mathcal{P}}(p_T,r,\phi)$ is obtained from the above equation dividing by the total number of particles produced, $N$, which in turn is obtained integrating over all phase space. Therefore
\bea
   {\mathcal{P}}(p_T,r,\phi)=\frac{1}{N}\frac{dN}{p_Tdp_Td^2r}.
\label{probdensity}
\eea

\begin{widetext}
\begin{figure*}[t]
\begin{center}
\includegraphics[scale=0.42]{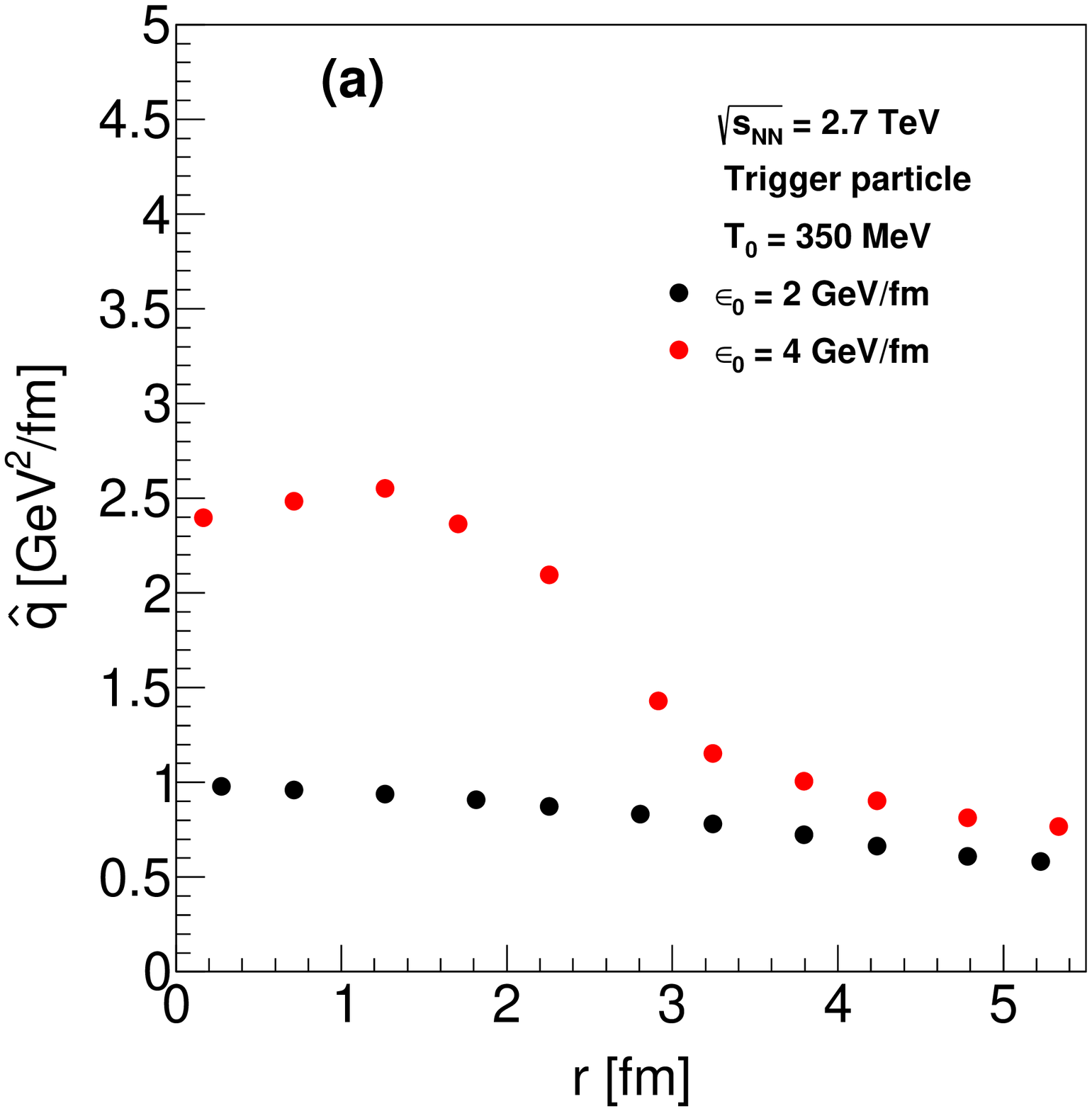}
\includegraphics[scale=0.42]{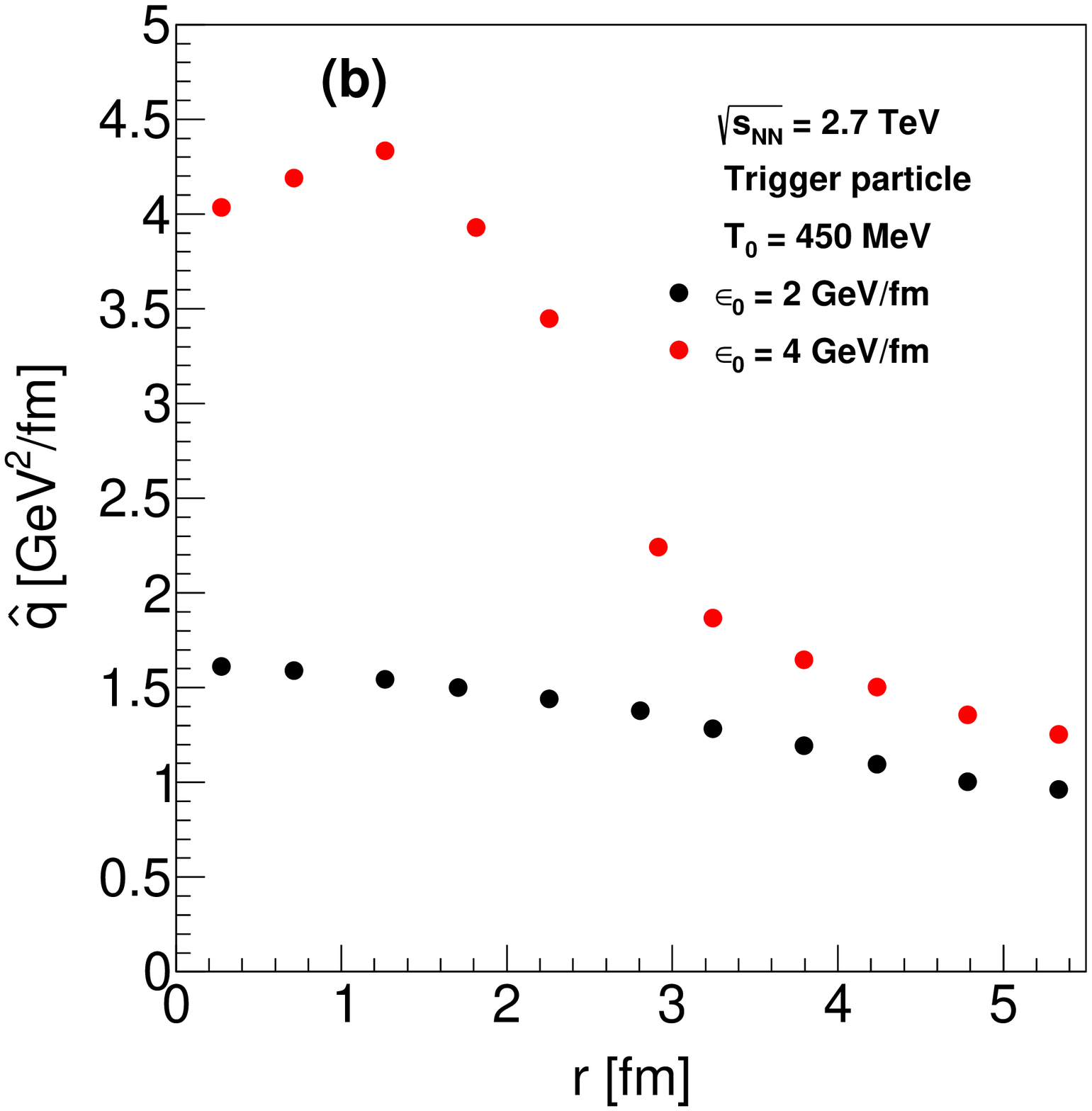}
\end{center}
\caption{$\hat{q}$ for the trigger particle as a function of the distance $r$ where the hard scattering took place. The left (right) panel corresponds to a QGP temperature  $T_0=350$ ($T_0=450$) MeV for the two values of $\epsilon_0=2,\ 4$) GeV/fm in the $1-d$ energy loss model.}
\label{fig2}
\end{figure*}

\begin{figure*}[t]
\begin{center}
\includegraphics[scale=0.42]{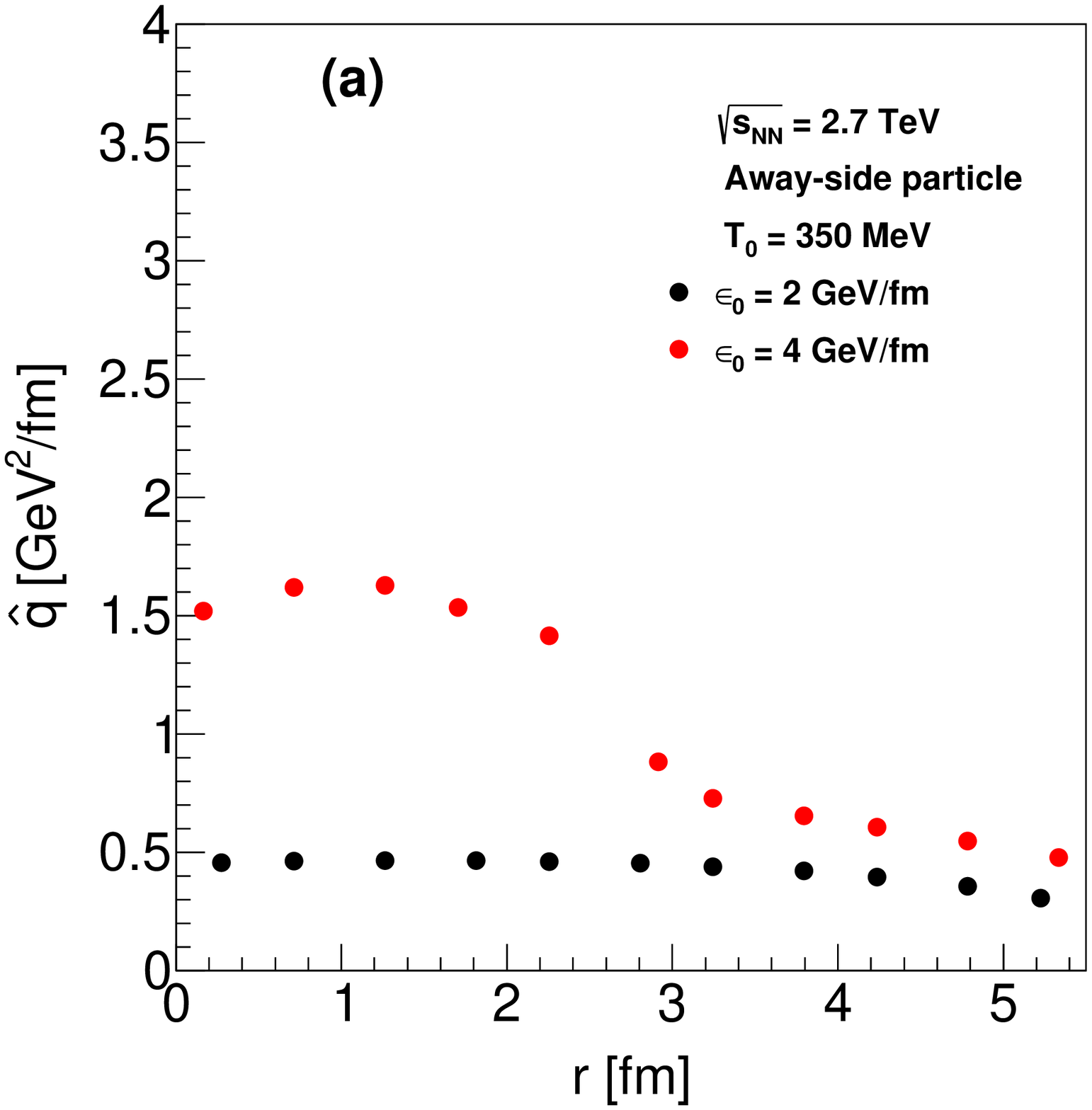}
\includegraphics[scale=0.42]{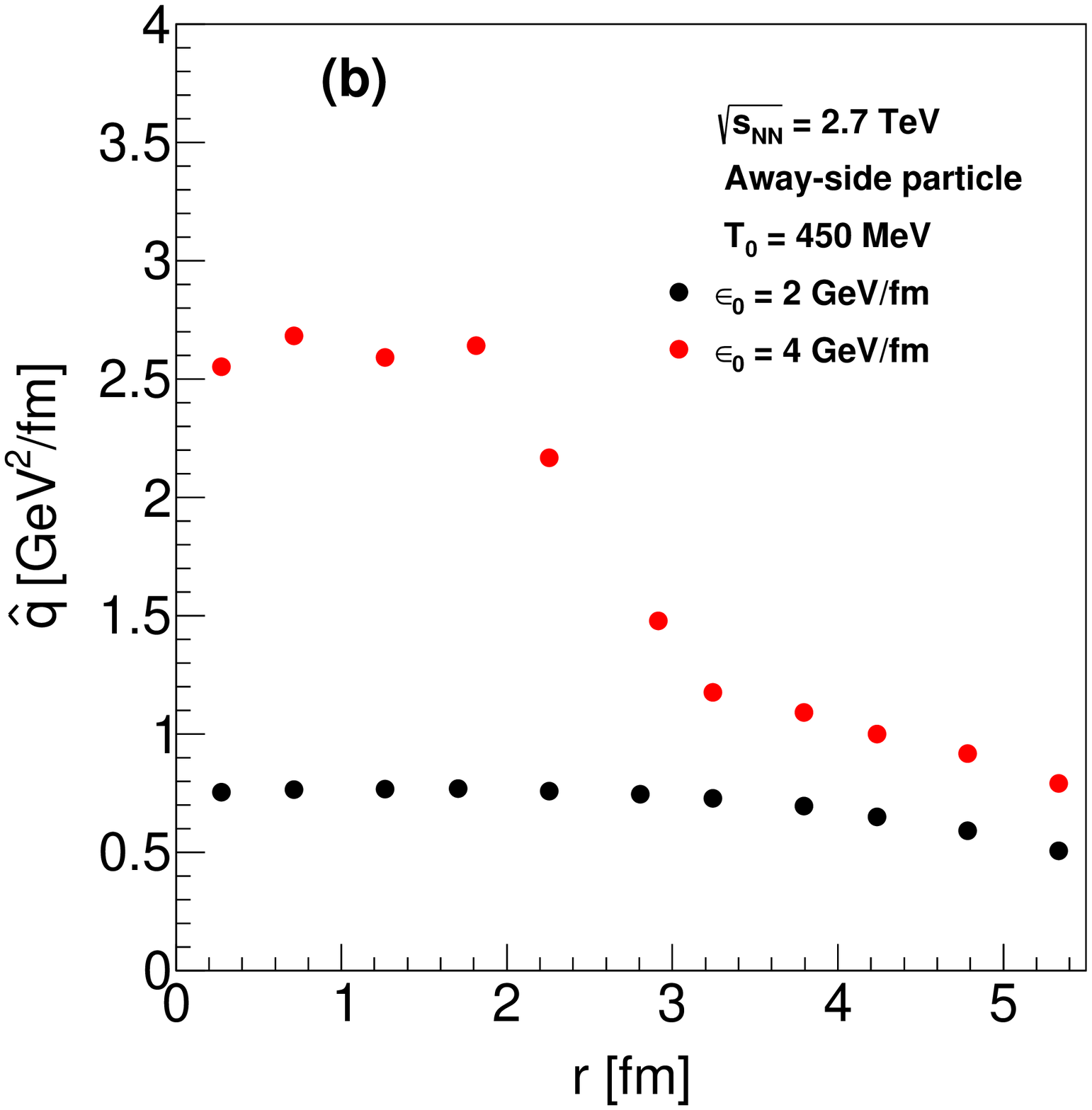}
\end{center}
\caption{$\hat{q}$ for the away-side particle as a function of the distance $r$ where the hard scattering took place. The left (right) panel corresponds to a QGP temperature  $T_0=350$ ($T_0=450$) MeV and a corresponding $\epsilon_0=2$ ($\epsilon_0=4$) GeV/fm in the $1-d$ energy loss model.}
\label{fig3}
\end{figure*}
\end{widetext}

Note that the average momentum squared carried by the disturbance, transverse to the direction of the fast parton's motion is given by
\bea
   \langle q^2 \rangle &\equiv& \int  d^2r \int dp_Tp_T\nn
   &\times&2\int_0^{\pi/2} d\phi\ {\mathcal{P}}(p_T,r,\phi)p_T^2\sin^2\phi,
\label{avtransmom}
\eea 
where in the integration over the angle between the fast parton and the produced particles we have implemented the condition to consider that these last move in the forward direction, namely, the direction of the fast parton.

Using Eqs.~(\ref{lineardist}) and~(\ref{probdensity}), $ \langle q^2 \rangle$ can be explicitly written as
\bea
   \langle q^2 \rangle &=& \frac{1}{N}\frac{\Delta\tau(\Delta y)^2}{(2\pi )^3}\frac{120T_0^5}{\epsilon_0}\nn
   &\times&
   \int d^2r \left[ \frac{\pi}{8}\delta\epsilon  + \frac{(4/3){\mathbf{g}}_y + (2/3){\mathbf{g}}_z}{(1+c_s^2)} \right].
\label{explicitqsquared}
\eea
Since $N$ is in turn given by
\bea
   N &=& \int  d^2r \int dp_Tp_T\nn
   &\times&2\int_0^{\pi/2} d\phi\ {\mathcal{P}}(p_T,r,\phi)\nn
   &=&\frac{\Delta\tau(\Delta y)^2}{(2\pi )^3}\frac{6T_0^3}{\epsilon_0}   
   \int d^2r \left[ \frac{\pi}{4}\delta\epsilon  + \frac{2{\mathbf{g}}_y + 2{\mathbf{g}}_z}{(1+c_s^2)} \right],
\label{N}
\eea 
we finally get
\bea
   \langle q^2 \rangle &=& 20\ T_0^2\
   \frac{\int d^2r \left[ \frac{\pi}{8}\delta\epsilon  + \frac{(4/3){\mathbf{g}}_y + (2/3){\mathbf{g}}_z}{(1+c_s^2)} \right]}
          {\int d^2r \left[ \frac{\pi}{4}\delta\epsilon  + \frac{2{\mathbf{g}}_y + 2{\mathbf{g}}_z}{(1+c_s^2)} \right]}.
\label{q-hatexpl}
\eea

Can $\langle q^2 \rangle$ be identified with the average momentum squared {\it given to the fast parton} by the medium and therefore with $\hat{q}$ upon dividing by the medium's length? The question is pertinent in the sense that the above calculation refers to the average momentum squared  {\it given to the medium by the fast parton}. If the parton's change in energy is small
the main effect on the fast parton is a deflection of its original trajectory. 
This deflection comes along with energy and momentum deposited within the medium via radiation or collisional processes.
As a result of energy and momentum conservation during these processes, the momentum put into the medium should compensate the momentum given to the fast parton. In other words, one could expect that the collected overall momentum (squared) should correspond to the equivalent quantity gained by fast parton in the transverse direction. Therefore, since in a hydrodynamical picture, the energy-momentum is described in terms of $\delta\epsilon$ and ${\mathbf{g}}$, we can write for the parameter $\hat{q}$
\bea
   \hat{q}=\frac{\langle q^2\rangle}{L},
\label{identify}
\eea
where $\langle q^2\rangle$ is given by Eq.~(\ref{q-hatexpl}).

Note that for the static case thus discussed, the expression for $\hat{q}$ is essentially independent of $\eta/s$. This happens because given the explicit factorization of $\Gamma_s$ and thus of $\eta/s$ in Eqs.~(\ref{moddeltaeps}) and~(\ref{modgtlyz}), the remaining dependence of $\eta/s$ cancels between numerator and denominator in Eq.~(\ref{q-hatexpl}) since the numerical coefficients accompanying $\delta\epsilon$ and the components of ${\mathbf{g}}$ are practically the same.

Since the assumption of an static medium is not entirely realistic, we now proceed to study whether an expanding medium makes $\hat{q}$ to depend on $\eta/s$.

\section{$\hat{q}$ and $\eta/s$ in an expanding medium}\label{III}

A full-fledged hydrodynamical computation of $\hat{q}$ in an expanding medium requires a numerical treatment. Let us instead attempt a phenomenological description based on modelling the way the medium gets diluted during the first stages of the collision due to longitudinal expansion~\cite{Wang}. A fast moving parton looses energy depending on the evolving gluon density that it traverses along its path through the medium, such that
\bea  
   \Delta E &=& \Bigl\langle \frac{d E}{d x}\Big\rangle_{1d} \, \int_{\tau_0}^{\infty} d
   \tau \frac{\tau - \tau_0}{\tau_0\, \rho_0}\, 
   \rho_g (\tau, \mathbf{b}, \mathbf{r} + \mathbf{\hat{n}} \tau),
\label{eq:delE}
\eea
where the gluon density $\rho_g$ is related to the nuclear geometry of the produced medium as
\bea
   \!\!\!\!\rho_g (\tau,\mathbf{b}, \mathbf{r}, \mathbf{\hat{n}}) &=&  \frac{\tau_0\,
   \rho_0}{\tau}\, \frac{\pi R_A^2}{2 A}\nn
   \!\!\!\!&\times& \left[ 
   T_A (|\mathbf{r} + \mathbf{\hat{n}} \tau |)
   +T_A (| \mathbf{b} - \mathbf{r} - \mathbf{\hat{n}}
   \tau |)\right]\!\!.
\label{meddens}
\eea

\begin{widetext}
\begin{figure*}[t]
\begin{center}
\includegraphics[scale=0.42]{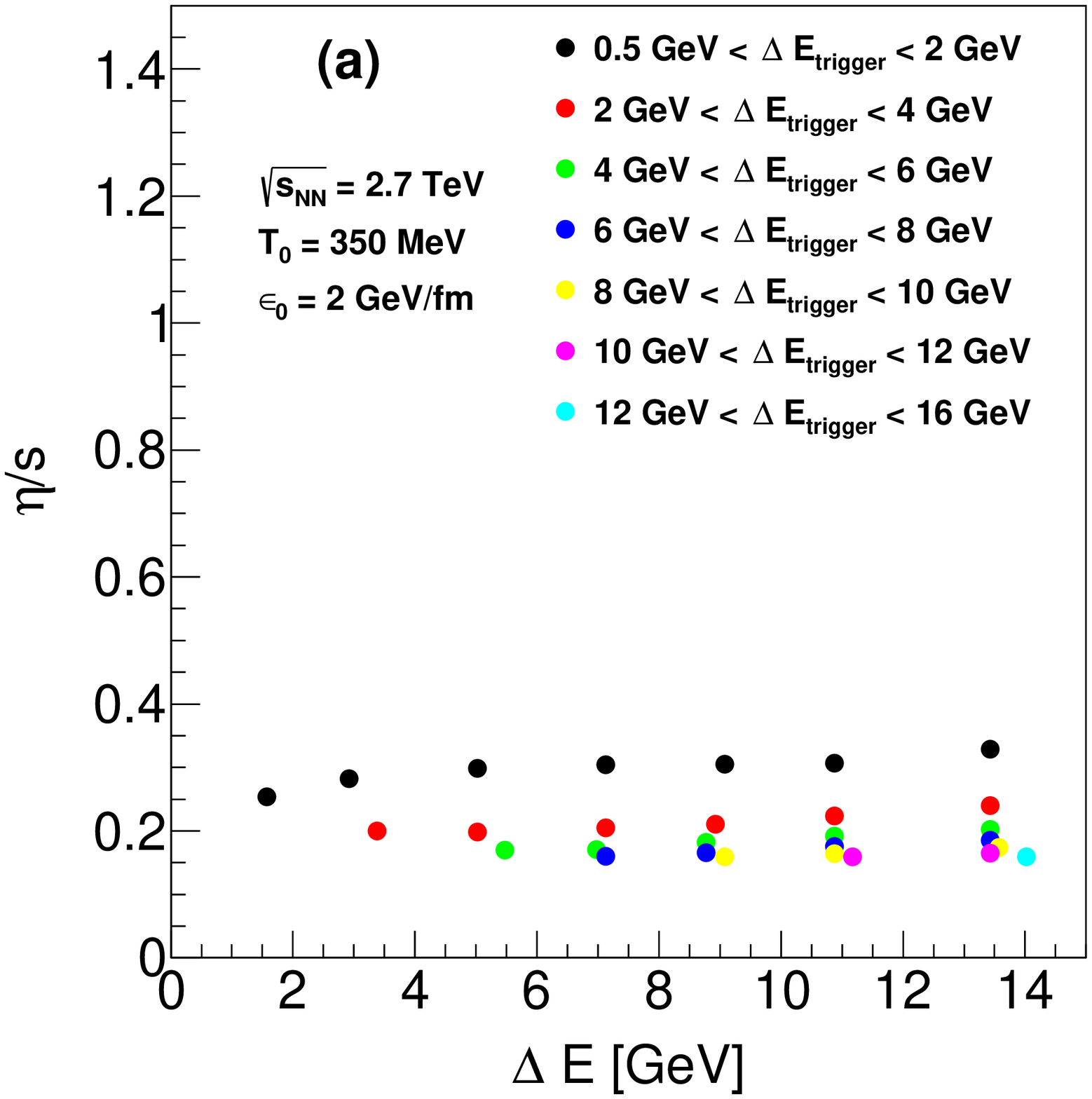}
\includegraphics[scale=0.42]{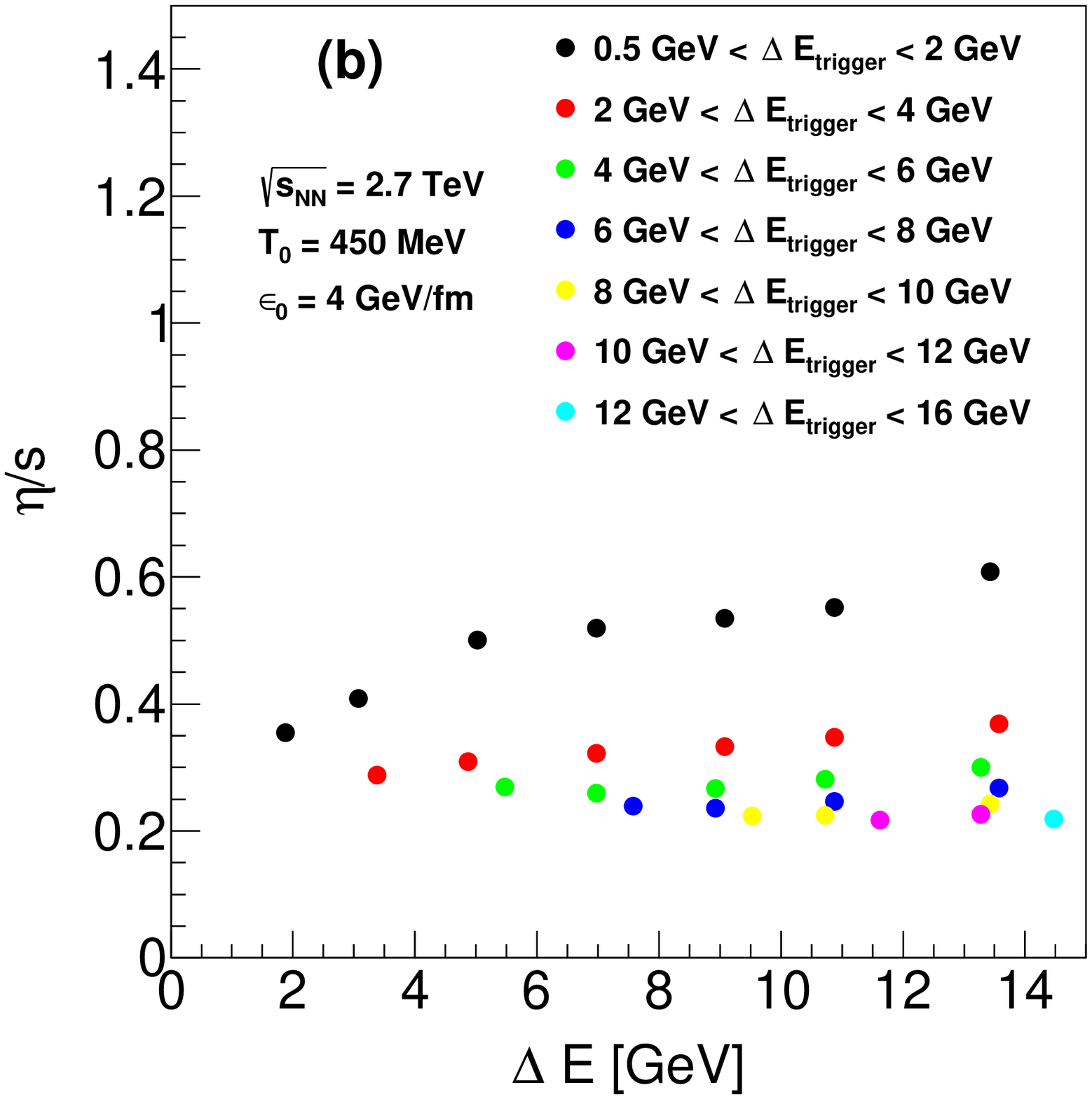}
\end{center}
\caption{$\eta/s$ as a function of $\Delta E$ for the away-side particle for different trigger particle energy loses. The left (right) panel corresponds to a QGP temperature  $T_0=350$ ($T_0=450$) MeV and a corresponding $\epsilon_0=2$ ($\epsilon_0=4$) GeV/fm in the $1-d$ energy loss model.}
\label{fig4}
\end{figure*}
\end{widetext}

Here $\rho_0$ is the central gluon density, $T_A$ the nuclear thickness function, $R_A$ the nuclear radius and $A$ the atomic number, $\mathbf{b}$ the impact parameter of the collision, $\mathbf{r}$ the transverse plane location of the hard scattering where the partons are produced and $\mathbf{\hat{n}}$ the direction in which the produced hard parton travels in the medium. The average number of scatterings $\langle n\rangle$ is given in the model by
\bea
   \langle n\rangle &=& \int_{\tau_0}^{\infty}  d
   \tau\frac{1}{\lambda_0\, \rho_0}\, 
   \rho_g (\tau, \mathbf{b}, \mathbf{r}, \mathbf{\hat{n}}\tau),
\label{avenumscatt}
\eea
where $\lambda_0$ is the parton mean free path for a constant density $\rho_0$. Since we want to consider the
most central collisions, hereafter we set $\mathbf{b}=0$. The one dimensional energy loss $\langle d E/d x\rangle_{1d} $ is parametrized as  
\bea
   \Bigl\langle \frac{d E}{d x}\Bigr\rangle_{1d} &=& \epsilon_0
   \Bigl[\frac{E}{\mu_0} - 
    1.6\Bigr]^{1.2}\Bigl[7.5 + \frac{E}{\mu_0}\Bigr]^{-1},
\label{dEdL}
\eea
where $E$ is the energy of the fast moving parton.  The mean free path for a constant density $\rho_0$ is taken as $\lambda_0=0.25$ fm. The parameter $\epsilon_0$ is related to $\lambda_0$ by $\epsilon_0\lambda_0=0.5$ GeV. We work with a value $\mu_0=1.5$ GeV. These parameters are tuned to describe LHC data on $R_{AA}$~\cite{Liu}.

In order to incorporate an expanding medium into the hydrodynamical description of the computation of $\hat{q}$ we approximate the average energy loss per unit length that appears in Eqs.~(\ref{moddeltaeps}) and~(\ref{modgtlyz}) with the energy loss given by the above described model divided by the in-medium length $L$ travelled by the fast moving parton, namely
\begin{eqnarray}
   \left( \frac{dE}{dx} \right) &=&\frac{\Delta E}{L(\mathbf{r},\mathbf{\hat{n}})},
   \label{approx}
\end{eqnarray}
where $\Delta E$ is given by Eq.~(\ref{eq:delE}). Note that with this choice the current $J^{\nu}(\mathbf{x},t)$ in Eq.~(\ref{current}) is still constant in space-time, however its amplitude depends on the parton's energy and on the matter density in the expanding medium. We take for $L(\mathbf{r},\mathbf{\hat{n}})$ the geometrical distance from the point where the hard scattering took place to the sharp edge of the interaction region, namely
\bea
   L(\mathbf{r},\mathbf{\hat{n}})=\frac{1}{2}\left(\sqrt{R_A^2-r^2\sin\varphi}-r\cos\varphi\right),
\label{Lsharp}
\eea
where $r=|\mathbf{r}|$ and $\varphi$ is the angle between $\mathbf{r}$ and $\mathbf{\hat{n}}$.  With this choice we account for the fact that within a diluting medium the mean free path $L(\mathbf{r},\mathbf{\hat{n}})/\langle n\rangle$ becomes larger than in the static case.

\begin{widetext}
\begin{figure*}[t]
\begin{center}
\includegraphics[scale=0.42]{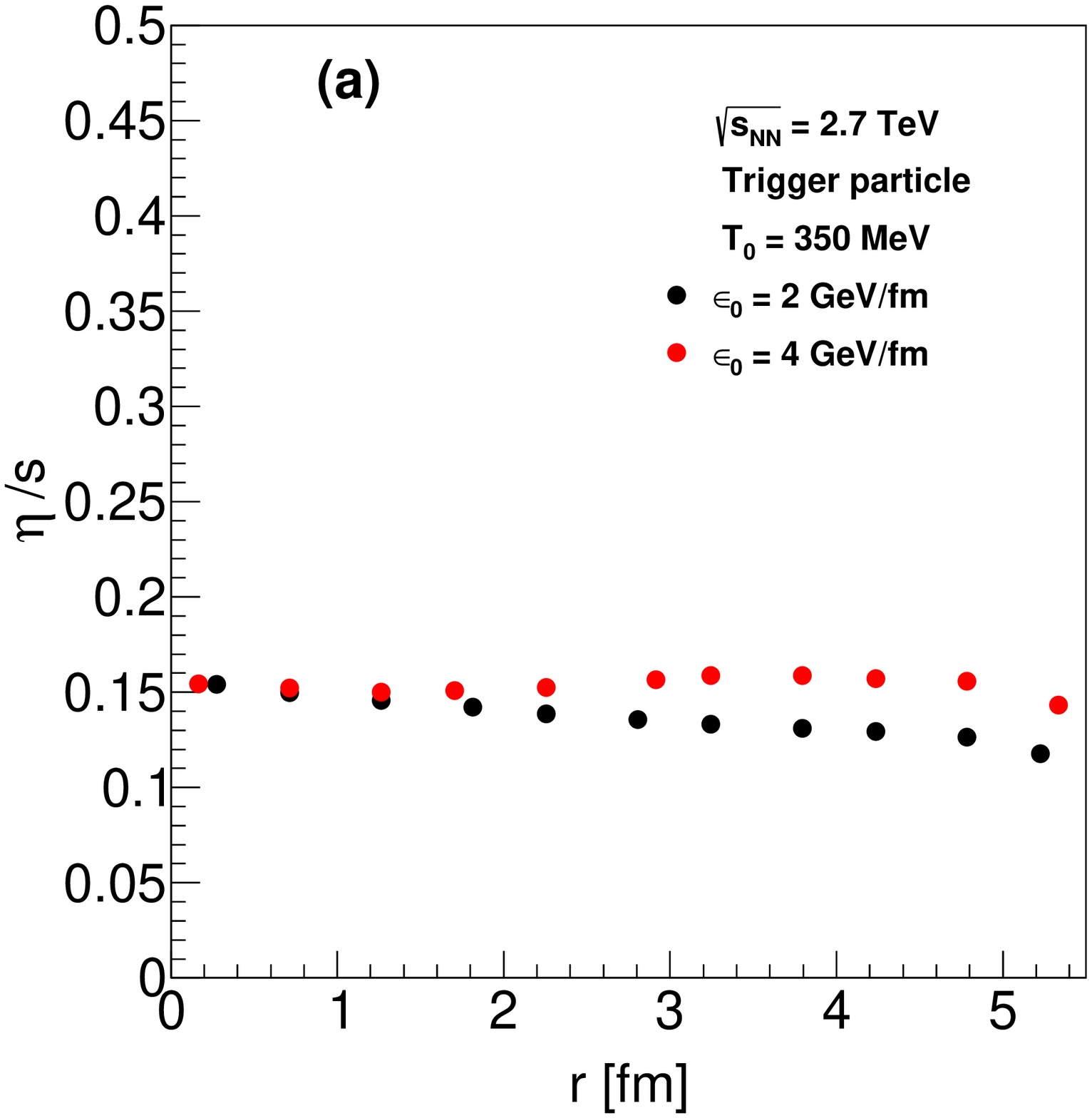}
\includegraphics[scale=0.42]{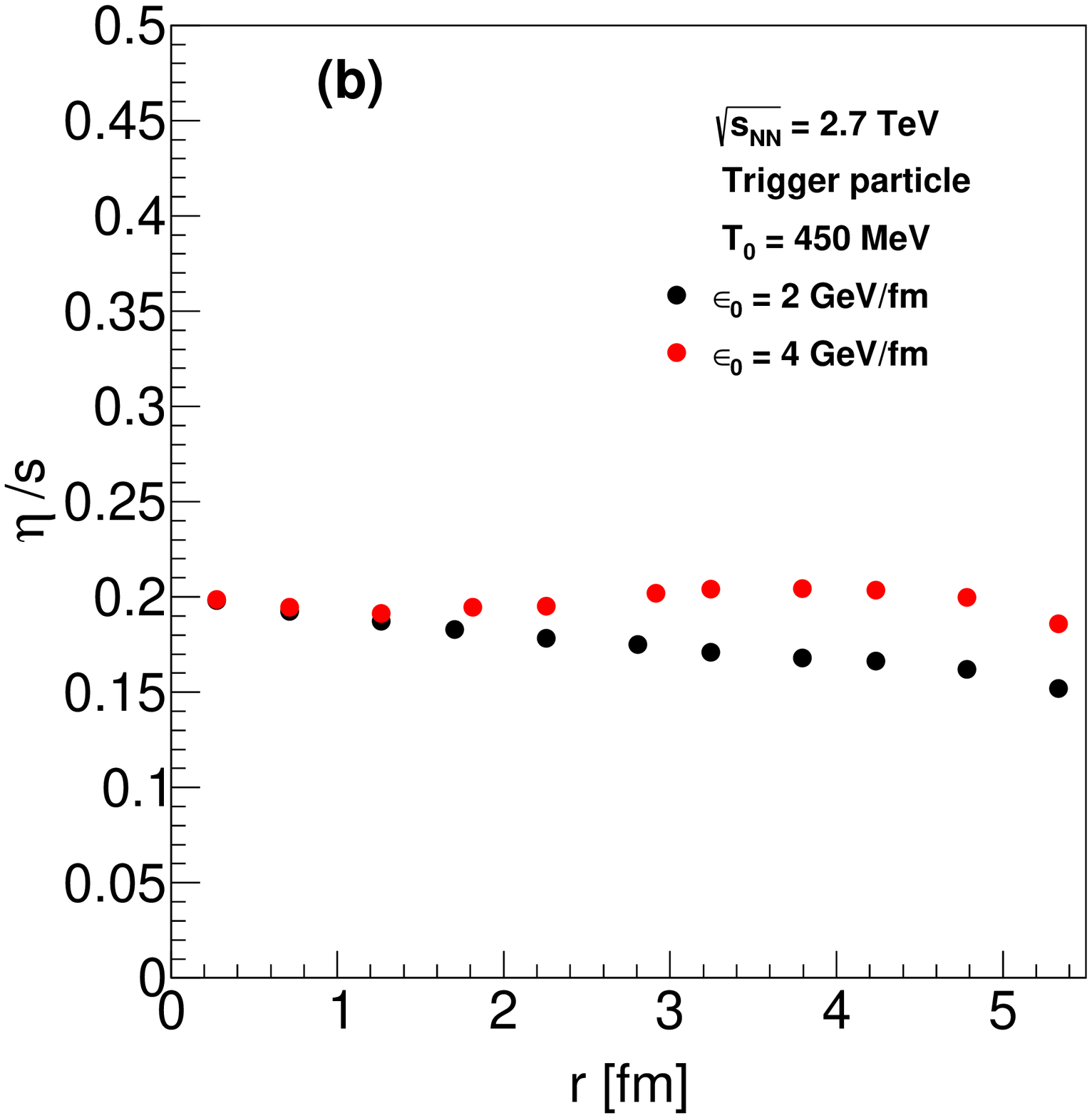}
\end{center}
\caption{$\eta/s$ for the trigger particle as a function of the distance $r$ where the hard scattering took place. The left (right) panel corresponds to a QGP temperature  $T_0=350$ ($T_0=450$) MeV and a corresponding $\epsilon_0=2$ ($\epsilon_0=4$) GeV/fm in the $1-d$ energy loss model.}
\label{fig5}
\end{figure*}

\begin{figure*}[t]
\begin{center}
\includegraphics[scale=0.42]{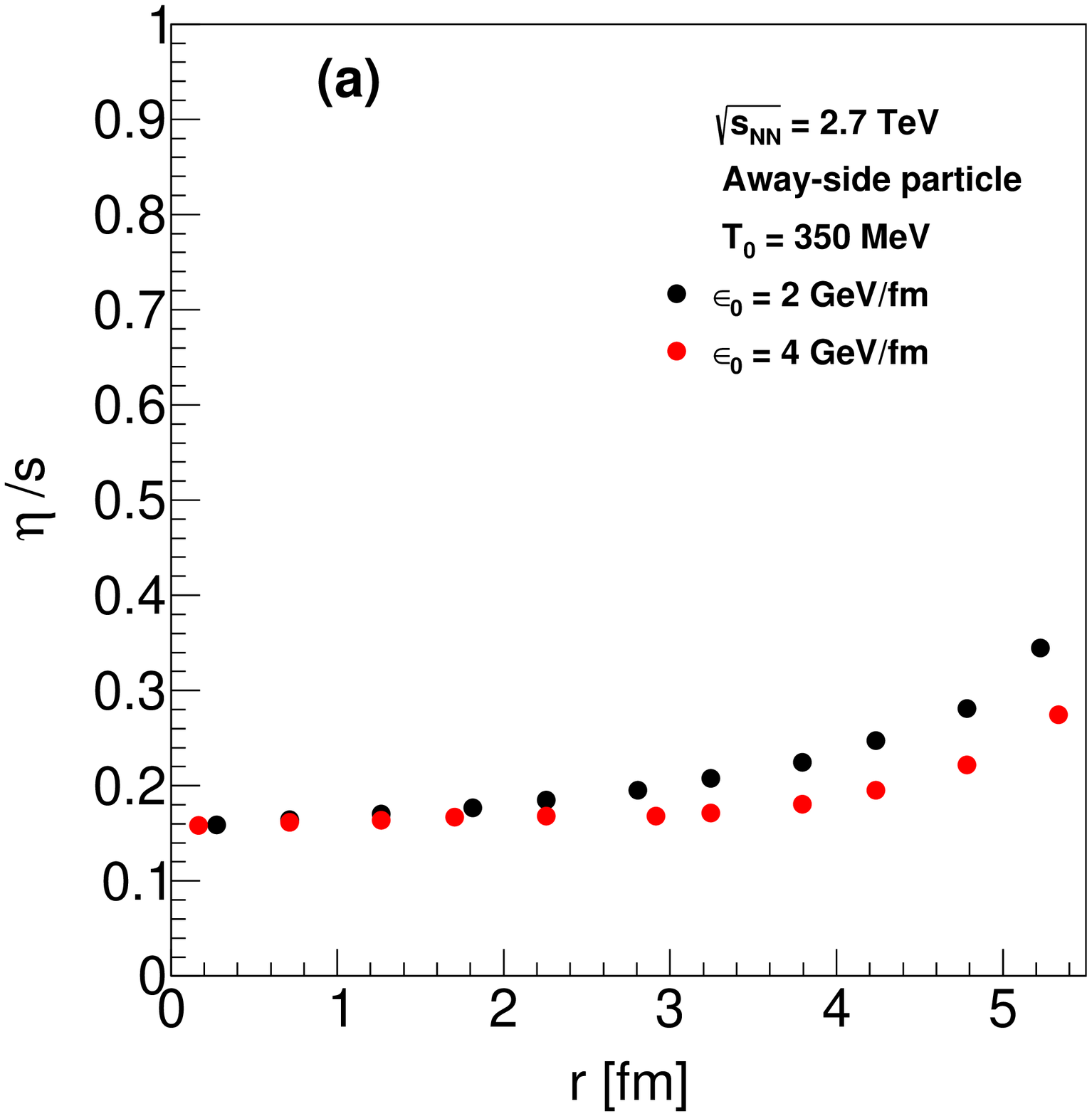}
\includegraphics[scale=0.42]{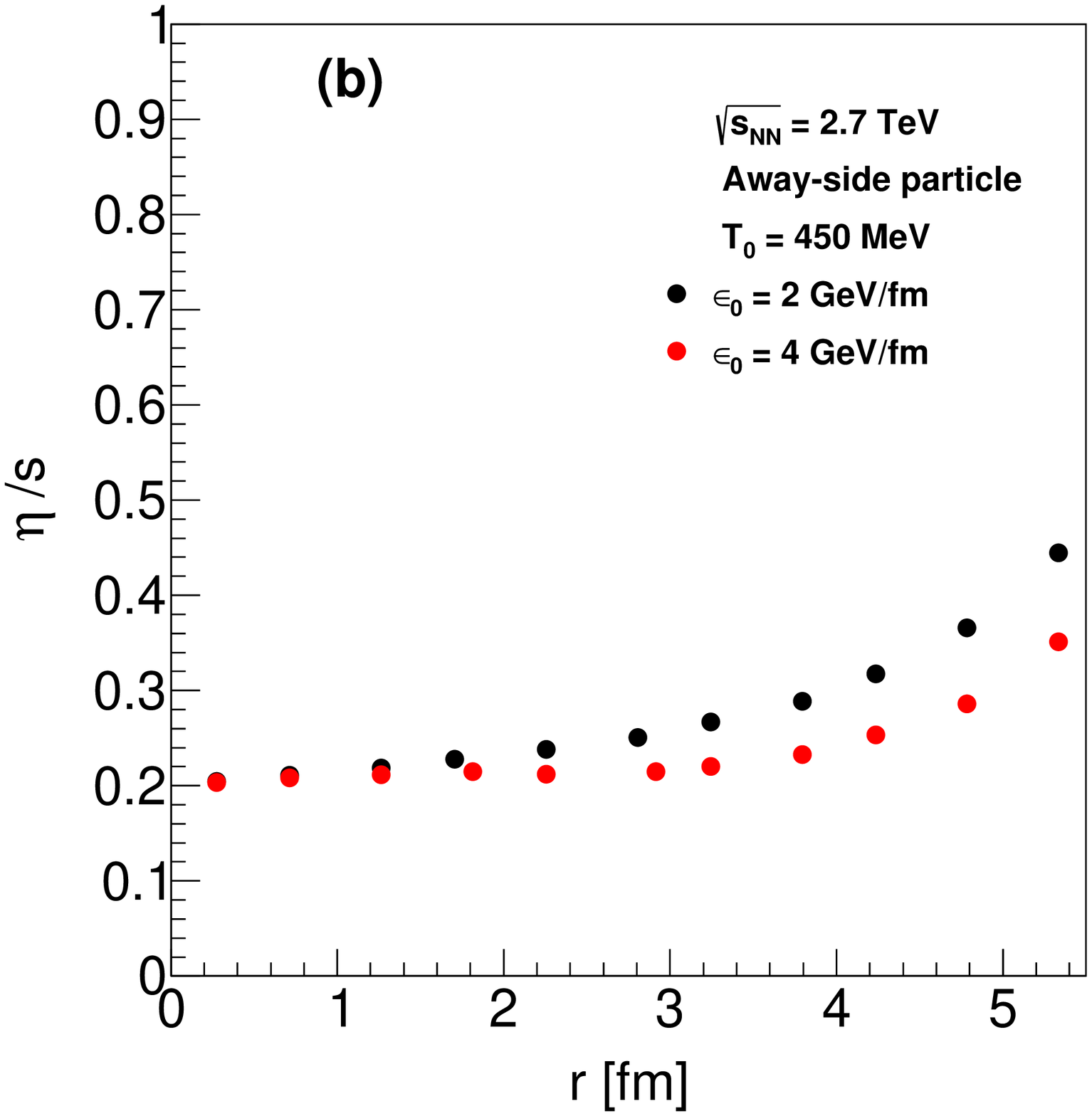}
\end{center}
\caption{$\eta/s$ for the away-side particle as a function of the distance $r$ where the hard scattering took place. The left (right) panel corresponds to a QGP temperature  $T_0=350$ ($T_0=450$) MeV and a corresponding $\epsilon_0=2$ ($\epsilon_0=4$) GeV/fm in the $1-d$ energy loss model.}
\label{fig6}
\end{figure*}
\end{widetext}

After these changes, Eqs.~(\ref{moddeltaeps}) and~(\ref{modgtlyz}) become
\begin{eqnarray}
  \delta\epsilon &=&
   \left(\frac{1}{4\pi}\right)
   \left( \frac{\Delta E}{L(\mathbf{r},\mathbf{\hat{n}})} \right)\left(\frac{2v}{3\Gamma_s}\right)^2 
   \left(\frac{9}{8v}\right)I_{\delta\epsilon}(\alpha ,\beta )
\label{moddeltaeps-2}
\end{eqnarray}
and
\begin{eqnarray}
\bd{g}_i &=&
   \left(\frac{1}{4\pi}\right)
   \left( \frac{\Delta E}{L(\mathbf{r},\mathbf{\hat{n}})} \right)\left(\frac{2v}{3\Gamma_s}\right)^2 I_{\mathbf{g}_i}
   (\alpha ,\beta ),
\label{modgtlyz-2} 
\end{eqnarray}

To compute $\hat{q}$ we now generate a sample of parton events at random positions $\mathbf{r}$, moving in random directions $\mathbf{\hat{n}}$ within the medium.  This sample is obtained using MadGraph 5~\cite{madgraph} for $2\to 2$ parton events in p + p collisions at $\sqrt{s_{NN}}=2.7$ TeV that subsequently loose energy according to the model thus described. In this manner we produce a distribution of events characterized by values of $\Delta E$, $L(\mathbf{r},\mathbf{\hat{n}})$ and $\langle n\rangle$. Therefore $\hat{q}$ can be obtained for instance as a function of $\Delta E$ by classifying events with a given amount of energy lost, regardless of where the hard scattering took place or the direction of motion of the fast parton, as
\begin{eqnarray}
   \hat{q}_{\Delta E}&=&
   \frac{20\ T_0^2\int d^2r \left( \frac{\pi}{8}\delta\epsilon  + \frac{(4/3){\mathbf{g}}_y + (2/3){\mathbf{g}}_z}
   {(1+c_s^2)}
   \right)_{\Delta E}}
   {\sum_{\Delta E}L(\mathbf{r},\mathbf{\hat{n}})\int d^2r \left( \frac{\pi}{4}\delta\epsilon  + \frac{2{\mathbf{g}}_y + 2{\mathbf{g}}_z}
   {(1+c_s^2)} \right)_{\Delta E}}.
\label{q-hatbin}
\end{eqnarray}
Alternatively,  $\hat{q}$ can also be obtained as a function of the location of the hard scattering, regardless of the direction the parton travelled or the amount of energy lost. 

\begin{widetext}
\begin{figure*}
\begin{center}
\includegraphics[scale=0.42]{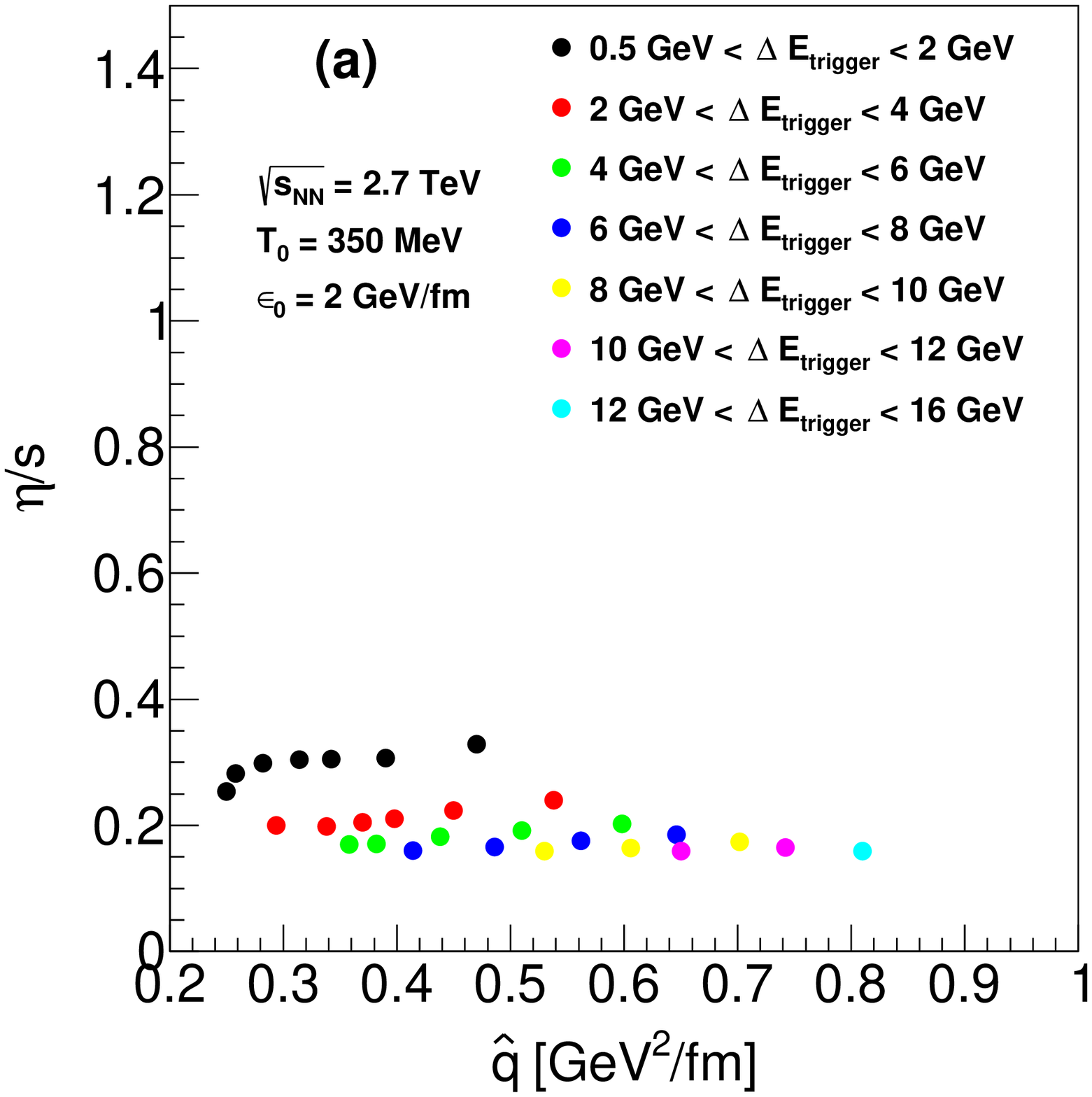}
\includegraphics[scale=0.42]{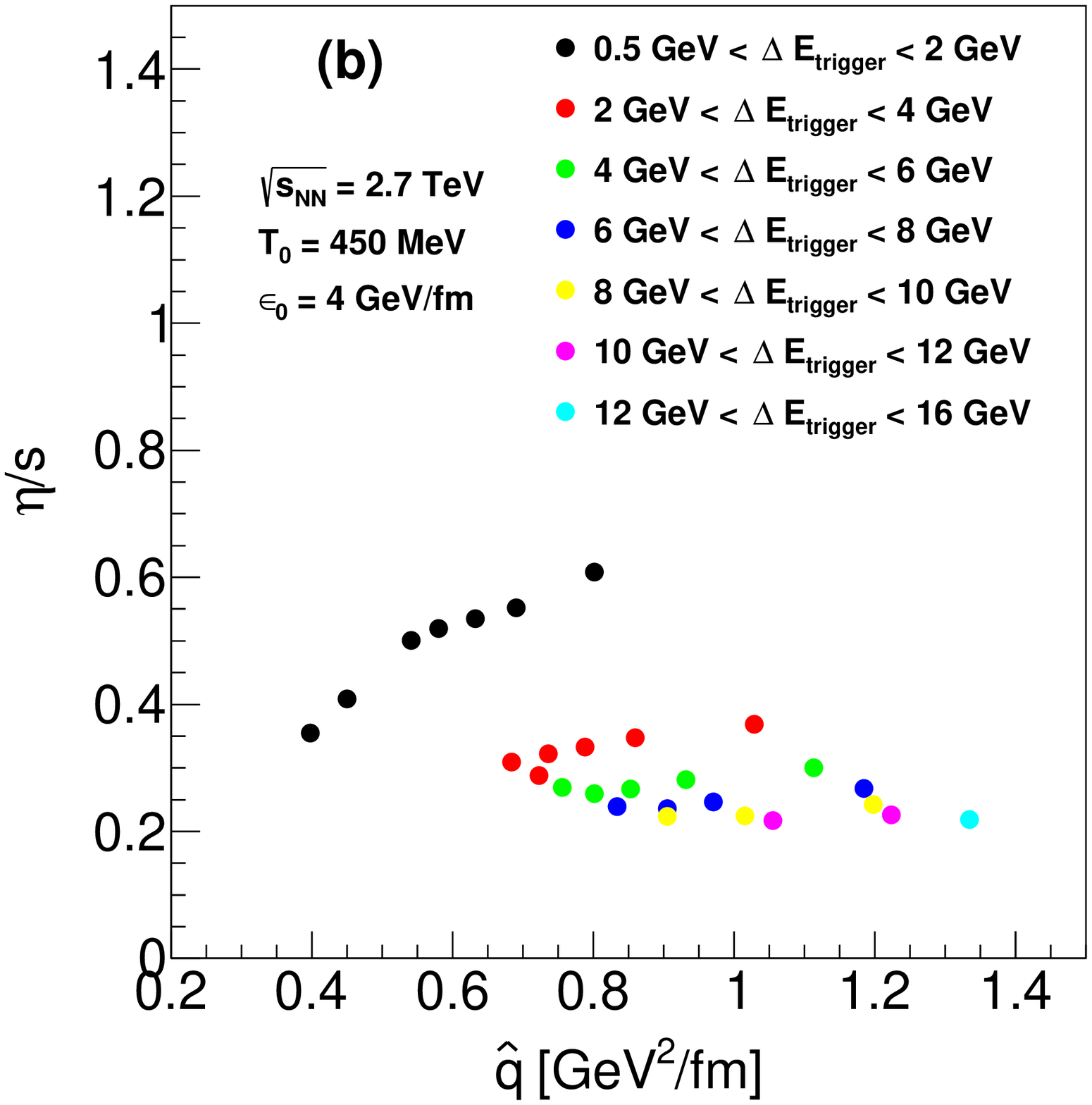}
\end{center}
\caption{Relation between $\eta/s$ and $\hat{q}$ for the away-side particle, for different trigger particle energy loses. The relation is obtained by plotting the value of these parameters that correspond to a given value of $\Delta E$.  The left (right) panel corresponds to a QGP temperature  $T_0=350$ ($T_0=450$) MeV and a corresponding $\epsilon_0=2$ ($\epsilon_0=4$) GeV/fm in the $1-d$ energy loss model.}
\label{fig7}
\end{figure*}
\end{widetext}

Notice that the model can also be used to estimate $\eta/s$ since we know that this transport coefficient is proportional to the ratio of the medium's mean free path to the thermal wavelength~\cite{Heinz,GyulassyQGP}. The model gives, event by event
\bea
   \frac{\eta}{s}\sim T\ \frac{L(\mathbf{r},\mathbf{\hat{n}})}{\langle n\rangle},
\label{etaovers}
\eea
where $\langle n\rangle$ is given by Eq.~(\ref{avenumscatt}). Again, we can classify events with a given energy loss, also regardless of where the hard scattering took place or the direction of motion of the fast parton or as a function of the location of the hard scattering, regardless of the direction the parton travelled or the amount of energy lost. In this way we can correlate events with a total amount of energy loss to the corresponding value of $\eta/s$ which characterizes the amount of medium traversed by the fast parton. 

Since the total energy lost by the fast parton bears a relation with the amount of medium the parton traversed, we can also study the transport coefficients classifying the events in terms of the value of $r$ where they were produced. On general grounds one expects that, modulo the influence of the expanding medium, partons that travel a larger path length loose more energy.

Figure~\ref{fig1} shows $\hat{q}$ as a function of the energy lost by the away-side particle for the cases where the trigger particle loses the indicated amount of energy. Hereby, for this and the rest of the figures, the left (right) panel corresponds to a QGP temperature  $T_0=350$ ($T_0=450$) MeV computed with $\epsilon_0=2,4$ GeV/fm in the $1-d$ energy loss model. Notice that the $\hat{q}$ values are widespread and have a strong dependence on the amount of energy lost both by the trigger and the away-side particles, making it difficult to assign a unique value of this parameter to characterize the plasma. 

To study a possible geometrical effect caused by the amount of effective medium travelled by the partons on the value of $\hat{q}$, Figs.~\ref{fig2} and~\ref{fig3} show $\hat{q}$ for the trigger and associate particles, respectively, as a function of the distance $r$ where the hard scattering took place measured from the center of the interaction region.  Notice that $\hat{q}$ has a similar behaviour for trigger and associate particles. There is a strong dependence of the value of $\hat{q}$ on $r$. The largest value occurs at a distance that maximizes the medium density as this last expands. Notice that contrary to the case when $\epsilon_0=2$ GeV/fm, the case $\epsilon_0=4$ GeV/fm shows a pronounced maximum for $\hat{q}$ around $r=1.4$ fm. This behavior can be understood by realizing that for the considered energy in the collision $\sqrt{s_{NN}}=2.7$ TeV, the average transverse momentum of the produced partons in the hard scattering is of order 15 GeV. For $\epsilon_0$ of order 1-2 GeV/fm partons lose on average 3-5 GeV which is small compared to the 15 GeV that partons were produced with on average. Therefore, since 1-2 GeV is not the patron's most likely value, the behavior of $\hat{q}$ is pretty much flat as a function of $r$, which is tantamount to $\Delta E$ given that partons lose more energy the more medium they travel. On the other hand, for $\epsilon_0$ of order 4 GeV/fm, partons lose on average 10 GeV which is
a quantity comparable to the average momentum they were produced with. Since $\hat{q}$ is basically the average momentum (squared) given from the partons to the medium and partons that travelled more medium (small to intermediate $r$) lose more energy, then, for bins with those values of $r$ there should be a maximum in
$\hat{q}$ since on average the majority of partons are produced with that value of momentum.

We can also study the behaviour of $\eta/s$ either as a function of the energy lost or as a function of the amount of effective medium travelled by the partons. Figure~\ref{fig4} shows $\eta/s$ as a function of $\Delta E$ of the away-side particle for the cases where the trigger particle loses the indicated amount of energy.  Notice that as the trigger particle loses more energy, the away side particle energy loss starts at higher values, as expected.  $\eta/s$ is larger for the case where the trigger particle loses less energy. For all events but the ones where $0.5<\Delta E_{\mbox{\small{trigger}}}<2$ GeV the $\eta/s$ values cluster around 0.2 (0.3) for the left (right) panel. In this case we have checked the behaviour of $\Delta E$ is dominated by events where the trajectory is tangential to the medium's surface, where, though $L$ is small, the amount of traversed medium is also small and so is the average number of scatterings. In all cases $\eta/s$ reaches limiting values which depend on the amount of energy lost by the trigger particle. Though the spread in values is less marked than for the case of $\hat{q}$, there is still a mild dependence on the energy loss by the trigger particle.

Figures~\ref{fig5} and~\ref{fig6} show $\eta/s$ as a function of $r$ for the trigger and associate particles, respectively. Notice that the value of this transport coefficient is more or less constant as a function of the location of the scattering center, where the fast parton is created. This can be understood from recalling that from Eq.~(\ref{etaovers}) $\eta/s$ is the ratio of the medium's travelled length to the average number of scatterings, scaled by the medium's temperature. Since the average number of scatterings is proportional to the travelled length, $\eta/s$ is constant modulo the effect of the varying density as the medium expands. Also, $\eta/s$ decreases (increases) with $r$ for trigger (associate) particles. This can be understood as an effect caused by the amount of medium traversed by the corresponding particle. A trigger (associate) particle emitted close to the surface travels on average less (more) medium and this gets reflected in the behaviour of $\eta/s$ with $r$.

Figure~\ref{fig7} shows the relation between $\eta/s$ and $\hat{q}$ for the away-side particle, for different trigger particle energy loses. This relation can obtained from Eqs.~(\ref{q-hatbin}) and~(\ref{etaovers}) since for a given $\Delta E$ bin we have both a value of $\eta/s$ and of $\hat{q}$. As the trigger particle loses more energy, the away side particle energy loss starts at higher values. Also, for all events but the ones where $0.5<\Delta E_{\mbox{\small{trigger}}}<2$ GeV the $\eta/s$ values cluster around 0.2 (0.3) for the left (right) panel as $\hat{q}$ varies. Once again, the behaviour of $\Delta E$ in this mentioned case is dominated by events where the trajectory is tangential to the medium's surface, where both $L$ and $\langle n\rangle$ are small. We see that, though the dependence of $\eta/s$ on $\hat{q}$ is non-trivial due to effects caused by the expanding medium, different $\hat{q}$ values are described by more or less the same value of $\eta/s$. Overall, the milder dependence of $\eta/s$ on events with different $\Delta E$ or coming from different locations $r$ makes this transport coefficient to be a more accurate quantity to characterize the expanding plasma than $\hat{q}$.

\section{Summary and conclusions}\label{IV}

In conclusion we have shown that a non-trivial behaviour of the transport coefficients $\hat{q}$ and $\eta/s$ with the the location of the hard scattering and with the energy lost characterizing the events is obtained for an expanding medium. To obtain this behaviour we resorted to model the amount of energy and momentum given to the medium by a fast moving parton in terms of linear viscous hydrodynamics and the amount of particles produced by this energy-momentum in terms of the Cooper-Frye formula. This procedure allows to obtain a probability distribution to compute the transport coefficients. To include the effect of an expanding medium we resorted to the model advocated in Ref.~\cite{Wang} tuned to describe LHC data~\cite{Liu}. The fast moving partons are produced in $2\to 2 $ parton events in p + p collisions with $\sqrt{s_{NN}}=2.7$ TeV. This allows to characterized events where particles lose a given amount of energy or are produced at a given location within the medium.

The study shows that the expanding medium cannot be characterized by single values of $\hat{q}$ or $\eta/s$, though the second one of these coefficients shows a milder dependence on $r$ or $\Delta E$. These results show that for conditions present in nuclear collisions at high energies, it is important to characterize the events in terms of a given observable, such as the amount of energy loss (missing $p_t$), before extracting a particular value for the transport coefficients.

\section*{Acknowledgments}

Support for this work has been received in part from PAPIIT-UNAM under grant  number IN101515 and from {\it Programa de Intercambio UNAM-UNISON} and {\it Programa Anual de Cooperaci\'on Acad\'emica UAS-UNAM}. J. J-M. is supported
by the DOE Office of Nuclear Physics through Grant No. DE-FG02-09ER41620.

\end{document}